\documentclass[%
 reprint,
 amsmath,amssymb,
 aps,
]{revtex4-2}

\usepackage{graphicx}
\usepackage{dcolumn}
\usepackage{bm}
\usepackage{comment}

\usepackage{physics}
\begin{document}

\preprint{APS/123-QED}

\title{Programmable XY-type couplings through parallel spin-dependent forces on the same trapped ion motional modes
}

\author{Nikhil Kotibhaskar}
\author{Chung-You Shih}%
\author{Sainath Motlakunta}
\author{Anthony Vogliano}
\author{Lewis Hahn}
\author{Yu-Ting Chen}
\author{Rajibul Islam}
\affiliation{%
 Institute for Quantum Computing and Department of Physics and Astronomy, University of Waterloo, Waterloo, ON, N2L 3G1, Canada
}%

\date{\today}
\newcommand{\yb}{$^{171}\rm{Yb}^+\;$}
\newcommand{\upup}{$\ket{\uparrow \uparrow}$\;}
\newcommand{\downdown}{$\ket{\downarrow \downarrow}$\;}
\newcommand{\downup}{$\ket{\downarrow \uparrow}$\;}
\newcommand{\updown}{$\ket{\uparrow \downarrow}$\;}
\newcommand{\XX}{$\sigma_x^i \sigma_x^j \;$}
\newcommand{\YY}{$\sigma_y^i \sigma_y^j \;$}
\begin{abstract}
We propose and experimentally demonstrate an analog scheme for generating XY-type ($J_{ij}^x \sigma_x^i \sigma_x^j \;$ + $J_{ij}^y \sigma_y^i \sigma_y^j \;$) Hamiltonians on trapped ion spins with independent control over the $J_{ij}^x$ and $J_{ij}^y$ terms.
The Ising-type interactions $\sigma_x^i \sigma_x^j \;$ and $\sigma_y^i \sigma_y^j \;$ are simultaneously generated by employing two spin-dependent forces operating in parallel on the same set of normal modes.
We analytically calculate the region of validity of this scheme, and provide numerical and experimental validation with $^{171}\rm{Yb}^+\;$ ions.
This scheme inherits the programmability and scalability of the Ising-type interactions with trapped ions that have been explored in numerous quantum simulation experiments.
Our approach extends the capabilities of existing trapped ion quantum simulators to access a large class of spin Hamiltonians relevant for exploring exotic quantum phases such as superfluidity and spin liquids.
\end{abstract}

\maketitle

Trapped ions are ideal quantum simulators of interacting spin systems \cite{Monroe_2021_Programmable,Blatt2012} due to their tunable long-range interactions \cite{Garcia-Ripol_2005_Coherent,Kim_2009_Entanglement,Molmer_1999_Multiparticle}, long coherence times \cite{Wang_2021_Single} and high fidelity quantum state preparation and measurement \cite{Harty_2014_High-Fidelity}.
Interacting spin models illuminate a large variety of many-body phenomena such as quantum magnetism and phase transitions, spin liquids, superconductivity, and superfluidity \cite{Auerbach_1994_Interacting,sachdev_2011_Quantum}.
Spins encoded in the internal degrees of freedom (such as hyperfine states) of individual trapped ions can interact via off-resonant excitation of their collective phonon modes through laser-driven spin-dependent dipole forces.
By varying the laser parameters, long-range and tunable Ising-type interactions have been experimentally demonstrated and used in a large number of quantum simulations to explore both equilibrium and dynamic phenomena.
Further, it has been proposed that the inherent long-range Coulomb interactions can be used to realize an arbitrarily programmable, all-to-all connected Ising spin system \cite{Korenblit_2012_Quantum,Teoh_2020_Machine}.

Existing proposals to simulate models that capture symmetries and phenomena beyond the reach of Ising models, such as XY and Heisenberg models, are either experimentally challenging or unfeasible, or limited in their applicability and tunability.
For example, non-local propagation of spin correlations \cite{Richerme_2014_Non-local,Jurcevic_2014_Quassiparticle} and many-body localization \cite{Smith2016_Many-Body} were studied, on an effective XY-Hamiltonian, by applying a large transverse magnetic field to the Ising Hamiltonian.
The transverse field restricts the Hilbert space contributing to the spin dynamics and results in an effective XY Hamiltonian only at discrete times \cite{Richerme_2014_Non-local, Kiely_2018_Relationship}.
The above approach also breaks down for long evolution times \cite{Kiely_2018_Relationship}, and does not allow the simulation of the anisoptopic XY model (i.e., interactions of the form $J^x_{i,j}$\XX + $J^y_{i,j}$\YY with $J^x_{i,j} \neq J^y_{i,j}$, where $\sigma_{x(y)}^i$ are the usual spin-$1/2$ Pauli matrices).  

Alternate proposals make use of orthogonal sets \cite{Davoudi_2020_Towards} of phonon modes, with each set mediating an independent Ising term (such as \XX or \YY).
This requires selective excitation of phonon modes along different spatial directions and the beams that excite phonon modes along one trap axis,
must not excite the other orthogonal set of phonon modes.
However, exciting multiple sets of orthogonal  phonon modes `selectively' requires additional laser beams, electronic controls, and complex optical design beyond the scope of current experimental setups. 
Further, this scheme may not produce the same form and range of interactions along different spin axes \cite{Monroe_2021_Programmable}, limiting the usefulness of such simulations.

Here, we demonstrate the creation of XY-type interactions, including the anisotropic XY-model, that simulates the equivalent spin dynamics in continuous time (limited by the coherence time of the system), and is readily implemented in existing experimental setups.
Our key insight is that the same set of motional modes can mediate both \XX and \YY interactions and the error in the evolution can be made negligible with the proper choice of the applied spin-dependent forces.
We experimentally demonstrate the dynamics of two \yb ion spins under the simulated XY Hamiltonian and numerically show that the scheme is scalable for larger systems.

\begin{figure*}
\includegraphics[width =1.0\linewidth]{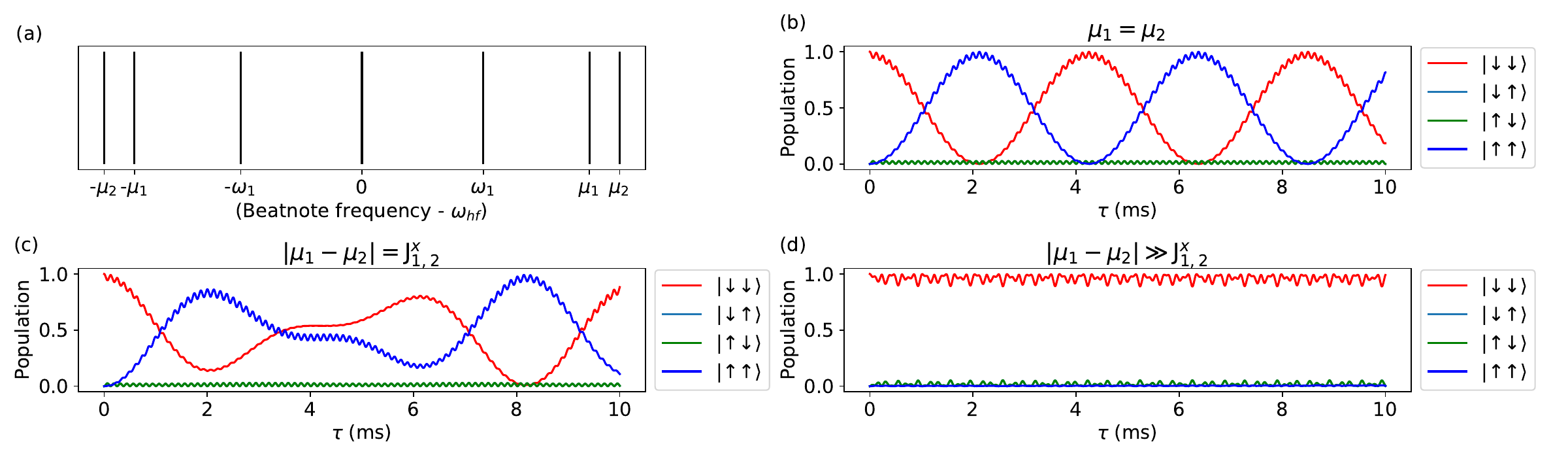}
\caption{\label{fig:schematic} 
\textbf{Schematic and numerical simulation of the protocol} }
\textbf{(a)} Schematic of the applied beatnotes generating SDFs at frequencies $\mu_1$ (for coupling to $\sigma_x$ spin components) and $\mu_2$ (for coupling to $\sigma_y$ spin components). 
The frequency axis is not drawn to scale.
The time evolution of the quantum state initialized in \downdown, shown in (b)-(d), are numerically calculated (using the python QuTiP library \cite{Johansson_2013_QuTiP}) from Eq. (\ref{eq_two_SDF}).
Probabilities of the four basis states (indicated in the plot legends) are calculated after tracing out the phonon degrees of freedom.
Here, $\omega_1 = 2\pi \times$ 1.1 MHz, $\mu_1 = 2\pi \times$ 1.108 MHz, $\psi_i^x=\psi_i^y=0$, $\eta_{i1}=0.0648$, and $\Omega_i^x = 2\pi \times$ 15 kHz ($i = 1, 2$). 
$\Omega_i^x$ is chosen to generate $J^x_{12} \approx 2\pi \times$ 60 Hz (calculated from Eq. (\ref{eq_2SDF_Jij}) ) and we truncate the motional Hilbert space to up to four phonons (five levels).
\textbf{(b)} Evolution for $\mu_1 = \mu_2$ and $\Omega_i^x = \Omega_i^y$ for  $i = 1, 2$.
Since we are effectively applying a single spin-dependent force, the interaction is simply Ising type (with the high-frequency oscillations, also seen in (c)-(d), arising from the off-resonant phonon excitation, i.e., from the first term in the exponent of Eq. (\ref{eq_Magnus_formula}) ).
\textbf{(c) }For $|\mu_1 - \mu_2| \approx J^{x}_{12}$ and $\Omega_i^x = \Omega_i^y$ for  $i = 1, 2$, the resulting dynamics is not pure XY-type due to the last two terms in the exponent of Eq. (\ref{eq_2SDF_U}) being non-negligible.
\textbf{(d)} For $|\mu_1 - \mu_2| \gg J^{x}_{12}$ ($\mu_2 = 2\pi \times$ 1.105 MHz), an effective XY-type Hamiltonian (Eq. (\ref{eq_main_result}) ) is realized. 
We set $\Omega_i^y = 2 \pi \times $ 12.2 kHz for  $i = 1, 2$ to set  $J^{x}_{12} = J^{y}_{12}$ and observe dynamics of the effective XY Hamiltonian (i.e., lack of oscillations between \downdown and \upup).
In choosing the exact value of $\Omega_i^y$ here, we use Eq. (\ref{eq_2SDF_Jij}), with an added correction (about $3\%$) to account for the AC Stark shift arising due to off-resonant excitation of the phonon mode, not considered in Eq. (\ref{eq_2SDF_U}).
 \end{figure*}

 \begin{figure*}
\includegraphics[width =1.0\linewidth]{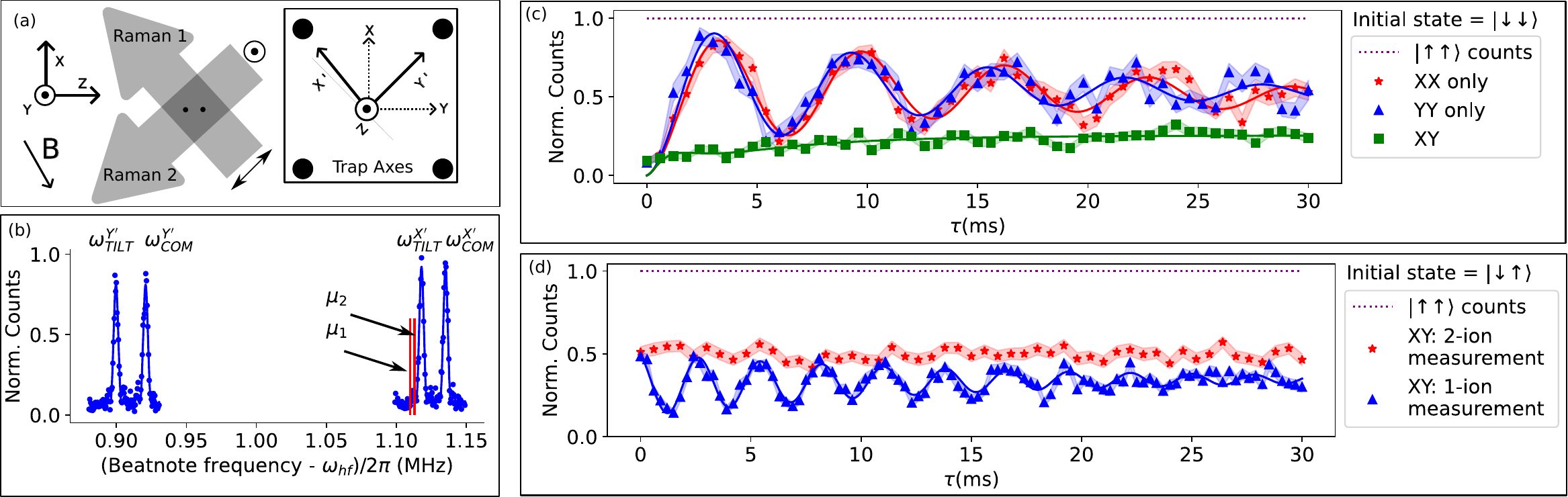}
\caption{ \label{fig:experiment_figure} 
\textbf{Experimental observation of dynamics under the XY Hamiltonian.} }
\textbf{(a)} Relative orientations of 2 raman beams w.r.t. ions (black dots), magnetic field, and beam polarization. 
The inset shows the transverse trap axes, $X'$ and $Y'$ in the XY plane with the circles representing the electrode (rod) positions.
The k-vector difference $\delta\vec{k}$ between the Raman beams lies along the X-axis and hence the Raman beams can excite both $X'$ and $Y'$ motional modes.
Experimental data are presented in (b)-(d), where the fluorescence counts are normalized w.r.t. the counts from the \upup state in each experimental repetition and each point is the average of 100 experimental repetitions. 
\textbf{(b)} Raman beatnote frequency scan with 2 ions showing the blue sidebands.
The higher frequency modes correspond to the $X'$ direction.
We have, $\omega^{X'}_{COM }= 2\pi\times 1.135 \; \mathrm{MHZ}$,  $\omega^{X'}_{\mathrm{TILT}} = 2\pi\times 1.117 \; \mathrm{MHZ}$,  $\omega^{Y'}_{COM}= 2\pi\times 0.920 \; \mathrm{MHZ}$,   $\omega^{Y'}_{\mathrm{TILT}} = 2\pi\times 0.899\; \mathrm{MHZ}$, $\mu_1 = \omega^{X'}_{\mathrm{TILT}} - 2\pi\times 8 \; \mathrm{kHz}$, $\mu_2 = \omega^{X'}_{\mathrm{TILT}} - 2\pi\times 5 \; \mathrm{kHz}$.
For (c) and (d), solid lines represent fit to $f(\tau)_{T,\alpha,\beta,\phi,C} = \sin^2(\omega \tau+\phi)(\alpha e^{-\tau/T}) +\beta(1-e^{-\tau/T})+C$ and the shaded regions represent standard error.
$\Omega_i^x \approx 2\pi \times$ 15 kHz ($i = 1, 2$) and $\Omega_i^y \approx 2\pi \times$ 11.5 kHz ($i = 1, 2$).
\textbf{(c)} When initialized in \downdown, we observe oscillations in the fluorescence counts, when either of the two spin-dependent forces are applied separately (red and blue plots). 
From the fit parameter $\omega$, we estimate $J^{x}_{12} = 2\pi\times$ 77(2) Hz and $J^{y}_{12} = 2\pi\times$ 80(3) Hz. 
When the two forces are applied simultaneously (green plots), we observe no appreciable oscillations, as expected from the XY Hamiltonian (see text).
\textbf{(d)} Observation of dynamics under the XY Hamiltonian, when initialized in \downup.
The normalized total fluorescence counts from both ions are nearly constant over the evolution time, as the spin states of the ions oscillate out of phase with each other (between \downup and \updown). 
Oscillations are recovered with a 1-ion fluorescence measurement (see text).
We observe oscillations at $2\pi\times$ 178(2) Hz, which is consistent with the expected oscillation frequency of $J^{x}_{12} + J^{y}_{12}$, counting for $\approx$ 15\% drifts in the couplings over the timescale of the data run.
The drifts also lead to a reduction in contrast of the coherent oscillations and with stabilization of both the intensity ($\Omega^{x(y)}_i$) and trap frequency ($\omega^{X'}_{COM }$ and $\omega^{X'}_{TILT}$)we expect an order of magnitude improvement in the coherent oscillations.
\end{figure*}
 
Spin-spin interactions can be induced \cite{Monroe_2021_Programmable} between ions by employing spin-dependent forces (SDF) that off-resonantly excite their collective vibrational modes.
Such SDFs can be applied using Raman transitions from lasers that are far detuned from unwanted atomic excitation.
The M{\o}lmer-S{\o}rensen scheme \cite{Molmer_1999_Multiparticle} for generating Ising-type interactions uses an SDF at a frequency $\mu$ that is generated by simultaneously applying Raman `beatnote' frequencies $\omega_{\mathrm{hf}} \pm \mu$ (the so-called `blue' and `red' sidebands) \cite{Lee_2005_Phase,Kim_2009_Entanglement}.
Here, $\omega_{\mathrm{hf}}$ is the frequency splitting between the two spin states.

Under the rotating wave ($\omega_{\mathrm{hf}} \gg \mu$) and the Lamb-Dicke approximations \cite{Monroe_2021_Programmable,Lee_2005_Phase,Zhu_2006_Trapped},
the spin-phonon Hamiltonian is:
\begin{equation}
\label{eq_MS_interaction_Hamiltonian}
    H = \sum_{i}  \Omega_i \cos( \mu t + \psi_i ) ( \delta \vec{k} \cdot \vec{x}_i ) \sigma_{\theta_i}^i
\end{equation}
where,  $\delta \vec{k} \cdot \vec{x}_i = \sum_m \eta_{im} ( \hat{a}_m e^{-i \omega_m t } + \hat{a}_m^\dagger e^{ i \omega_m t })$.
Here, $\Omega_i$ is the Rabi frequency at the $i^{\mathrm{th}}$ ion position,  $\hat{a}_m$ and $\hat{a}_m^\dagger$ are the phonon annihilation and creation operators for the $m^{th}$ motional mode at frequency $\omega_m$. 
The Lamb-Dicke parameters $\eta_{im} = b_{im} |\delta \vec{k}| \sqrt{\hbar/2M\omega_m}$ include the normal mode transformation matrix element $b_{im}$ of the $i^{\mathrm{th}}$ ion and  $m^{\mathrm{th}}$ normal mode \cite{James_1998_Quantum}. 
Where $\sum_i |b_{im}|^2 = \sum_m |b_{im}|^2 = 1 $, $M$ is the mass of the ion and $\sigma_{\theta_i}^i = \sigma_x^i \cos\theta_i + \sigma_y^i \sin\theta_i$.
The spin phase $\theta_i$ and the motional phase $\psi_i$ are determined from the relative phases of the red and blue sidebands \cite{Monroe_2021_Programmable}.
The evolution operator for this Hamiltonian can be found using the Magnus expansion, which terminates after the second term,
\begin{equation} 
\begin{split}
\label{eq_Magnus_formula}
    U(\tau) \! &= \!\exp \left(-i \int_0^\tau \!\!\!dt H(t) \! - \!\frac{1}{2} \int_0^\tau \!\! dt_1 \! \int_0^{t_1} \!\!dt_2 [ H(t_1), H(t_2) ] \right) \\ 
    & = \exp \left( \sum_i \hat{\phi}_i(\tau)\sigma_{\theta_i}^i  
    + i \sum_{i<j} \chi_{i,j}(\tau) \sigma_{\theta_i}^i \sigma_{\theta_j}^j \right). 
\end{split}        
\end{equation}

In the `slow' regime ($|\mu - \omega_m| \gg \eta_{im}\Omega_i$), $\hat{\phi_i}(\tau)$ is negligible \cite{Kim_2009_Entanglement} and $\chi_{i,j}(\tau)$ in Eq. (\ref{eq_Magnus_formula}) is dominated by a `secular' term, which grows linearly with $t$, giving rise to an effective Hamiltonian,

\begin{equation}
\label{Jij formula}
    H_{\mathrm{eff}} = \hbar \sum_{i<j} J_{i,j} \sigma_{\theta_i}^i \sigma_{\theta_j}^j \\ 
\end{equation}
where, the Ising coupling,
\begin{equation*}
    J_{i,j} = \Omega_i \Omega_j \frac{\hbar ( \delta \vec{k} )^2}{2M} \sum_m \frac{ b_{im} b_{jm}}{\mu^2 - \omega_m^2}. \nonumber
\end{equation*}
Note that, the unitary evolution in Eq. (\ref{eq_Magnus_formula}) will, in general, include additional AC Stark shifts such as from off-resonant excitation of the `carrier' spin transition from the SDFs (which we did not include in Eq. (\ref{eq_MS_interaction_Hamiltonian}) and (\ref{eq_two_SDF}), as in Refs. \cite{Kim_2009_Entanglement,Monroe_2021_Programmable} for simplicity, but they must be accounted for in experiments). 

While multiple SDFs with different spin phases operating in parallel have been theoretically suggested in \cite{Grab_2014_Trapped}, experiments have mainly used parallel SDFs with same spin phase \cite{Shapira_2023_Quantum} or a time varying spin phase\cite{Kranzl_2023_Observation,Kranzl_2023_Experimental,shapira2023programmable} to perform quantum simulations.
Recent experiments have used SDFs along two orthogonal modes to generate parallel quantum gates \cite{Yingyue_2023_Pairwise}.
In our protocol, we apply two SDFs at frequencies $\mu_1$ and $\mu_2$ (Fig. 1(a)), with both of them exciting (off-resonantly) the same motional modes. 
We choose the red and blue sideband phases to generate a spin phase of 0 (corresponding to $\sigma_{\theta_i} = \sigma_x$ in Eq. (\ref{eq_MS_interaction_Hamiltonian}) ) for the first SDF and  $\pi/2$ (corresponding to $\sigma_{\theta_i} = \sigma_y$) for the second SDF.
The resulting spin-phonon Hamiltonian becomes,
\begin{equation}
\label{eq_two_SDF}
\begin{split}
    H & = H_{x} + H_{y}, \\
    \mathrm{where,}\\
      &H_{x} = \sum_{i,m}  \eta_{im}\Omega_i^x \cos( \mu_1 t + \psi^x_i ) ( \hat{a}_m e^{-i \omega_m t } + h.c.) \sigma_x^i, \\
      &H_{y} = \sum_{i,m}  \eta_{im}\Omega_i^y \cos( \mu_2 t + \psi^y_i ) ( \hat{a}_m e^{-i \omega_m t } + h.c.) \sigma_y^i.
\end{split}
\end{equation}

Here, $\Omega^x_i$ and $\Omega^y_i$ are  Rabi frequencies, and $\psi_i^x$, $\psi_i^y$ are motional phases corresponding to the 2 SDFs respectively.
Again, if we operate each of the two forces in the slow regime ($|\mu_1 - \omega_m| \gg \eta_{im}\Omega_i^x$,  $|\mu_2 - \omega_m| \gg \eta_{im}\Omega_i^y$ ), the first term in the Magnus expansion is negligible \cite{Kim_2009_Entanglement} and the evolution operator arising from the second term in the Magnus expansion becomes,

\begin{equation}
\label{eq_2SDF_U}
\begin{split}
    U(\tau) \! &= \!\exp \left( - \frac{1}{2} \int_0^\tau \!\! dt_1 \! \int_0^{t_1} \!\!dt_2 \left[ H(t_1), H(t_2) \right] \right)  \\
     &= \exp \Bigg( -i\tau \sum_{i<j} J_{ij}^x \sigma_x^i \sigma_x^j -i\tau\sum_{i<j}J_{ij}^y \sigma_y^i \sigma_y^j  \\
     &+ \quad\quad\quad \sum_{i<j}\Lambda_{ij}(\tau)\sigma_x^i \sigma_y^j  + \sum_i \hat{\zeta}_i(\tau) \sigma_z^i \Bigg).
\end{split}
\end{equation}
Where,
\begin{equation}
\label{eq_2SDF_Jij}
\begin{split}
    &J_{ij}^x = \Omega_i^x \Omega_j^x \frac{\hbar ( \delta \vec{k} )^2}{2M} \sum_m \frac{ b_{im} b_{jm}}{\mu_1^2 - \omega^2_m},\\
    &J_{ij}^y = \Omega_i^y \Omega_j^y \frac{\hbar ( \delta \Vec{k})^2}{2M} \sum_m \frac{ b_{im} b_{jm}}{\mu_2^2 - \omega^2_m}.
\end{split}
\end{equation}

The first two terms in the exponent in Eq. (\ref{eq_2SDF_U}) come from $ [ H_{x(y)}(t_1),H_{x(y)}(t_2) ]$, and result in the desired spin-spin interactions.
The last two terms come from the cross terms, i.e. $ [ H_{x(y)}(t_1),H_{y(x)}(t_2) ]$, and lead to an undesirable spin-phonon coupling.
For $\mu_1=\mu_2$, the single frequency SDF can be rewritten with a different spin phase, as can be seen from Eq. (\ref{eq_two_SDF}), and therefore the resulting effective spin-spin Hamiltonian is Ising type ($\sigma_{\theta}^i \sigma_{\theta}^j$) (Fig. \ref{fig:schematic}(b)).
For $\mu_1 \neq \mu_2$,  there is no secular term in $\Lambda_{ij}(\tau)$ and $\hat{\zeta}_i(\tau)$, however, they may have non-trivial oscillatory terms (see appendix). 
As $|\mu_1 - \mu_2|$ is increased beyond zero, the contribution of the oscillatory terms diminishes (Fig. \ref{fig:schematic}(c)), and the evolution becomes consistent with an effective XY-type Hamiltonian (Fig. 1(d)),

\begin{equation}
\label{eq_main_result}
H_{\mathrm{eff}} = \hbar \sum_{i<j} J_{ij}^x \sigma_x^i \sigma_x^j  + \hbar \sum_{i<j} J_{ij}^y \sigma_y^i \sigma_y^j,
\end{equation}
when,
\begin{equation}
\label{eq_main_result_constraint}
    |\mu_1-\mu_2|  \gg \max_{i,j}(| J_{ij}^x| ), \quad |\mu_1-\mu_2|  \gg \max_{i,j}( |J_{ij}^y| ).  
\end{equation}

In the following section, we provide experimental validation of the above analysis.

The experiments are performed on \yb ions in a four-rod Paul trap with trap frequencies  
 $\omega_X \approx$  2$\pi \times$ 1.135 MHz, $\omega_Y \approx$  2$\pi \times$ 0.920 MHz and $\omega_Z \approx$  2$\pi \times$ 201 kHz.
 The spins are encoded in the two hyperfine `clock' states, $S_{1/2}\ket{F=0,m_F=0}$ ($\ket{\downarrow}$) and $S_{1/2}\ket{F=1,m_F=0}$ ($\ket{\uparrow}$), of the \yb ions, separated in energy by the hyperfine splitting $\omega_{\mathrm{hf}}/2\pi$ = 12.643 GHz. 
Here, $F$ and $m_F$ are quantum numbers representing the total atomic angular momentum and its projection along a weak magnetic field of around 3.5 G.

We perform coherent operations on the ions, through 2-photon Raman transitions with a 355 nm pulsed laser with a repetition rate of ~80 MHz \cite{Islam_2014_Mode-locked}. 
The wave-vector difference of the two Raman beams, $\delta \vec{k}$, is oriented such that we can excite phonon modes along both transverse trap axes, $X'$ and $Y'$ (Fig. \ref{fig:experiment_figure}(a)) . 
We modulate the frequency of the Raman 1 beam with four harmonic tones to create four beatnotes driving SDFs at two frequencies $\mu_1$ =  $\omega^{X'}_{\mathrm{TILT}}$ - 2$\pi \times$ 8 kHz, $\mu_2$ = $\omega^{X'}_{\mathrm{TILT}}$ - 2$\pi \times$ 5 kHz respectively (Fig. \ref{fig:experiment_figure}(b)).
Here, $\omega^{X'}_{\mathrm{COM}}=\omega_X$ and  $\omega^{X'}_{\mathrm{TILT}}= 2\pi \times 1.117 \, \mathrm{MHz}$ are the frequencies of the COM and TILT modes in the $X'$ direction respectively.
The SDF detunings are chosen to be smaller than the separation between the modes to minimize the contribution from all modes except the $X'$ TILT mode.

The experimental sequence is as follows. 
We apply 1.5 ms of Doppler cooling and 8 ms of Raman sideband cooling to cool all transverse modes to $\Bar{n} < 1$ to be in the Lamb-Dicke regime, and global optical pumping for 20 $\mu$s to initialize in \downdown state. 
An optional $\pi$-pulse driven by microwave radiation (at frequency $\omega_{\rm{hf}}$) and a site-selective optical pumping (that maintains the coherence of the neighboring ion with $\sim 99.9\%$ fidelity \cite{Motlakunta_2023_Preserving}) can alternatively prepare the initial states \downup and \updown.
The spin-dependent forces are then applied (Eq. (\ref{eq_two_SDF})) for a pulse duration $\tau$.

We finally measure the spin states by state-dependent fluorescence on a photo-multiplier tube (PMT) for 1.5 ms.
We calibrate the fluorescence counts by preparing the spins in \upup with a microwave $\pi-$pulse in a separate experiment and obtain approximately 80 PMT counts for this state.
Since global fluorescence measurements cannot distinguish between \updown and \downup states, we apply a local optical pumping pulse on the first ion just before the measurement to convert it to a single ion measurement, when necessary.

\begin{figure}
\label{fig:Jij}
\includegraphics[width =\linewidth]{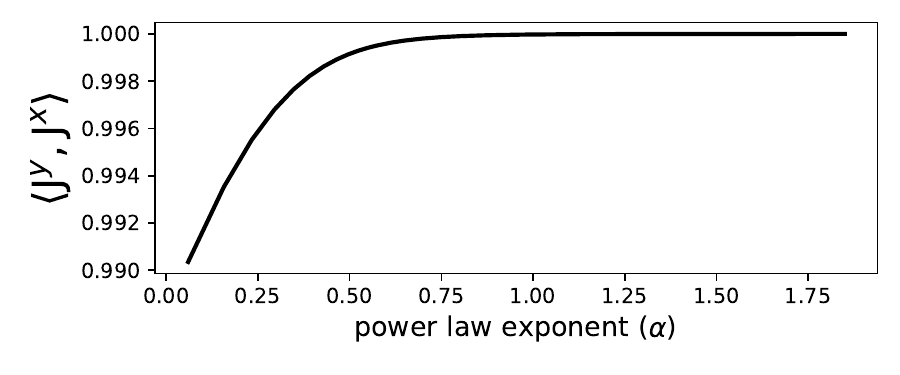}
\caption{
\textbf{Calculated proximity of $J^{x}$ and $J^{y}$ while engineering power law decay in XY couplings for $N=$ 25 ions.} 
To have a linear chain of $N=$ 25 ions, we assume trap frequencies of $\omega_X' =$  2$\pi \times$ 5 MHz, $\omega_Y' =$  2$\pi \times$ 4.8 MHz and $\omega_Z=$  2$\pi \times$ 400 kHz.
$J^{x}$ is calculated using Eq. (\ref{eq_main_result}), assuming equal $\Omega^x_i$ on all ions operating in the regime where $ \max(\{J_{ij}^x\}) \approx 2\pi \times$ 100 Hz.
$\mu_1$ is varied from $\omega_x + 2\pi \times$ 1kHz to $\omega_x+2\pi \times$  1 MHz to achieve the approximate power scaling, i.e., $J^x_{ij} \approx 1/|i-j|^\alpha$ with $\alpha\in (0.1,1.8)$.
To get the $J^x_{ij} = J^y_{ij}$, we set $\mu_2 = \mu_1+2\pi \times$ 3 kHz and calculate $\Omega_i^y$ from Eq. (\ref{eq_main_result}) by assuming only participation of the closest motional mode i.e., COM.
To compare $J^x$ and $J^y$ we use the Frobenius norm, $\langle \mathbf{A}, \mathbf{B} \rangle_\mathrm{F} = \mathrm{Tr}(\mathbf{A}^\dagger \mathbf{B} )$ after appropriate normalization.
}
\end{figure}

\begin{figure*}
\includegraphics[width =\linewidth]{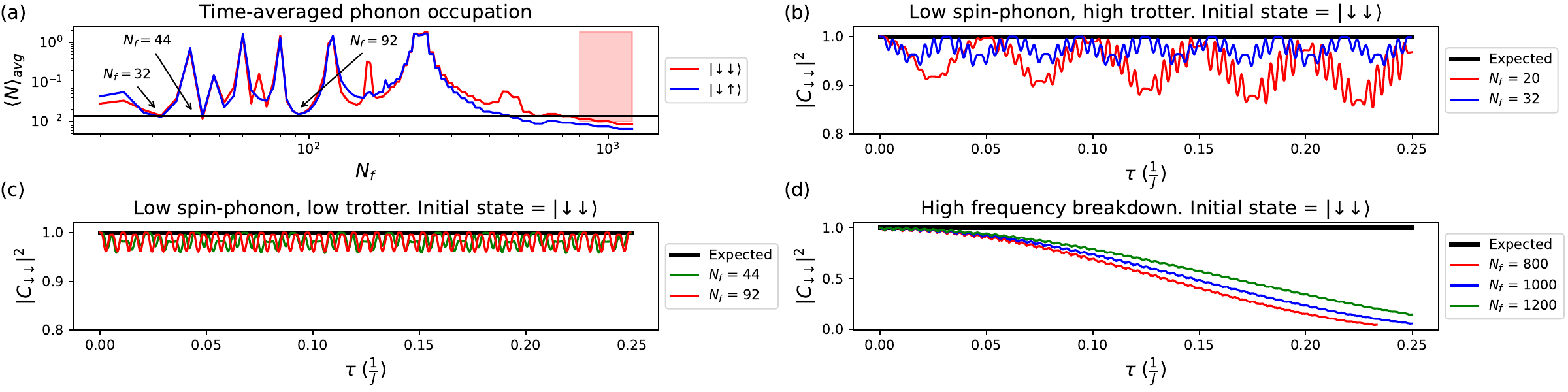}
\caption{ \label{fig:Floquet} \textbf{Comparison between Floquet engineering and analog generation of the XY Hamiltonian.} }
Numerically calculated time evolution of 2 ions initialized in the \downdown state under a Floquet pulse sequence intended to generate XY-couplings (see text).
We used a single motional mode to mediate the spin-spin interactions.
All the frequencies are scaled to this mode frequency, i.e.,  $\omega_m = 2\pi \times 1$, $\mu = 2\pi \times 1.02$ and $\eta \Omega = 2\pi \times 0.003$  (see Eq. \ref{eq_MS_interaction_Hamiltonian}) for both XX and YY pulses, intended to simulate the effective Hamiltonian $H_{eff} = J^x_{12} \sigma^1_x \sigma^2_x + J^y_{12} \sigma^1_y \sigma^2_y$  with $J^x_{12} = J^y_{12}  = J \approx 2\pi \times 8\times 10^{-5}$, and observe the dynamics in the timescale of $\tau = 1/J$.
We define $N_f = (\frac{1}{T_f \times J})$, where $T_f$ is the floquet period and find that resulting time evolution is dependent on $N_f$.
\textbf{(a)} Time averaged phonon population, ($\langle N \rangle_{avg}$) during the length of the quantum simulations when initialized in either \downdown or \downup. 
Most values of $N_f$ lead to inadvertent spin-motion coupling of the ions.
The ideal $N_f$ values are the ones where the average phonon number is comparable to the native `2-SDF' approach (black line) used in this work, for example, $N_f = $ 32, 44 and 92 here.
The population in the \downdown state as a function of time are shown for various values of $N_f$ in (b), (c) and (d).
\textbf{(b)} For low values of $N_f$, the trotter errors limit the efficacy of the simulation as evidenced by the slow oscillations at 1/$T_f$.
\textbf{(c)} The best values of $N_f$ are where the trotter error and the spin-motion coupling are both small.
\textbf{(d)} For sufficiently large values, $N_f \geq 800$ here (shaded red in (a)), the time evolution deviates from the expected significantly despite low phonon occupation.
\end{figure*}
Figure \ref{fig:experiment_figure}(c) shows the spin dynamics when initialized in \downdown state.
As in the numerical simulations Fig. \ref{fig:schematic}, we tune the Rabi frequencies to achieve $J^{x}_{12} \approx J^{y}_{12}$ to get an effective XY model when the interactions are mediated simultaneously.
We set the Rabi frequencies, $\Omega^x_i = 2 \pi \times$ 15 kHz and $\Omega^y_i = 2 \pi \times$ 11.5 kHz (approximately equal between the two ions).
With $H_x$ and $H_y$ applied separately, we observe oscillations between \downdown and \upup, as expected. 
We estimate $J^{x}_{12}$ = $2\pi \times\,$77(2) Hz and $J^{y}_{12}$ = $2\pi \times\,$80(3) Hz.
When applied simultaneously, we find no observable oscillations, which is the signature of the XY Hamiltonian, since ($\sigma_x^1 \sigma_x^2 + \sigma_y^1 \sigma_y^2$ )\downdown = 0. 
The slow increase in the fluorescence counts for the XY Hamiltonian is likely due to the decoherence in the system and slow drifts in the intensity of the laser and the trap frequency (estimated to contribute $<15\%$ drift in $J^{x}_{12}$ or $J^{y}_{12}$) over the duration of the data run.
To further validate that the \XX and the \YY couplings are present simultaneously, we initialize the ions in the \downup state and repeat the experiment.
Here, we expect oscillations between the \downup and the \updown states, at frequency $J^{x}_{12} + J^{y}_{12}$ (see appendix).
With a global detection, we expect the fluorescence counts to be flat as observed in Fig. \ref{fig:experiment_figure}(d).
However, with an individual detection on ion 1 (i.e. by pumping one ion before detection), we observe oscillations in the fluorescence counts as expected from oscillations between the \downup and \updown.
From the data in Fig. \ref{fig:experiment_figure}(d), we observe oscillations at $2\pi \times$ 178(2) which is within the expected 10\% fluctuations of the extracted $J^{x}_{12} + J^{y}_{12}$ from previous experiments.

The inherent full-connectivity of trapped ions allows for the scheme to be readily scalable to a large number of ions like in the case of the tunable Ising interactions \cite{Kim_2009_Entanglement,Monroe_2021_Programmable}.
For example, an interaction profile with a power law decay that has been widely studied for quantum Ising models \cite{Monroe_2021_Programmable} can be extended to XY interactions.
To achieve this interaction profile, first the Rabi frequencies ($\Omega_i^x$) and the detuning ($\mu_1$) can be chosen to obey an approximate power law in the coupling matrix $J^x$ in Eq. (\ref{eq_2SDF_Jij}).
Then, $\mu_2$ can be chosen to satisfy Eq. (\ref{eq_main_result_constraint}) while keeping it close to $\mu_1$, to maintain the same form for $J^x$ and $J^y$.
Further, if $J^x=J^y$ is desired, then $\Omega_i^y$s can be calculated using a global scaling w.r.t $\Omega_i^x$s to compensate for the unequal $\mu_1$ and $\mu_2$.
Figure 3 shows that the interaction profiles along $x$ and $y-$directions match to better than 99$\%$ with the global scaling of Rabi frequencies, even when $|\mu_1-\mu_2|$ is chosen to be 30 times higher than max($J^x$).
We find that this approach of scaling the Rabi frequencies works well whenever $\mu_{1}$($\mu_{2}$) is parked close to a motional mode since the contribution to the $J_{ij}^x$ ($J_{ij}^y$) is dominated by a single motional mode.

It shuold be noted that the constraint in Eq. (\ref{eq_main_result_constraint}) is weaker than the constraint for applying each of the forces in the slow regime ($|\mu_{1(2)} - \omega_m| \gg \eta_{im}\Omega_i^{x(y)}$ ) which is applicable even while applying a single SDF.
This is because, in the slow regime, individual matrix elements $J_{ij}^x$ ($J_{ij}^y$) are an order of magnitude smaller than $|\mu_{1(2)} - \omega_m|$ and hence leave enough freedom to satisfy Eq. (\ref{eq_main_result_constraint}) (See appendix for more details).
Thus, by simultaneously applying a pair of SDFs near each motional mode, the full spin-spin interaction profile can be engineered arbitrarily \cite{Korenblit_2012,Teoh_2020_Machine}.
It should be noted, however, that the calculation of the coupling matrix should take into account any AC Stark shift induced from the spin-dependent forces when the relative scale of the couplings needs to be precisely matched.

The XY-model (and the Heisenberg model) can also be implemented on trapped ions via Floquet engineering \cite{Kranzl_2023_Observation,Kranzl_2023_Experimental}.
Repeated application of a pulse sequence comprising of a \XX-SDF for time $T_f/2$ followed by a \YY-SDF with the same interaction profile for time $T_f/2$ leads to the XY-model in the high frequency limit ($T_f << 1/\max_{i,j}( |J_{ij}^{x(y)}| )$) \cite{Bukuv_2015_Universal}, where the Trotter errors are negligible.
Assuming no additional spin-motion coupling, varying the floquet period $T_f$ in this regime should leave the effective Hamiltonian unchanged except for the residual Trotter errors.
However, the floquet drives can be viewed as two separate SDFs that are further modulated at the floquet frequency (and harmonics) potentially leading to complicated dynamics and additional spin-motion coupling.
To asses these effects, we numerically simulate the evolution of 2 ions (and one motional mode) under such a sequence intended to simulate the effective hamiltonian, $H_{eff} = J\sigma^1_x \sigma^2_x + J\sigma^1_y \sigma^2_y$ (See Figure \ref{fig:Floquet} ).
We incorporate a Blackman pulse shape \cite{Kranzl_2023_Observation} at the beginning and end of the pulses with 40\% of the pulse time split evenly in the rising and falling edge to avoid broadening of the pulses in the frequency domain due to sharp rising/falling edges (see appendix).
We observe that the time evolution is highly dependent on $N_f=(\frac{1}{T_f \times J})$ and matches the expectation only at specific values of $N_f$ marked in Figure  \ref{fig:Floquet}(a).
We observe that the time-averaged phonon occupation serves as a good predictor for when the time evolution matches that of the target Hamiltonian, for $N_f\lesssim 800$.
For low $N_f$ values, Trotter errors are significant (Fig. \ref{fig:Floquet}(b)), even when phonon errors are low.
The `good' $N_f$ values (Fig. \ref{fig:Floquet}c), where both the Trotter and additional spin-motion coupling errors are low, depend on the Rabi frequency and the detuning applied to create the XX and YY pulses (see appendix).
Beyond $N_f\gtrsim 800$, the residual phonon occupation goes below the level expected from our dual-SDF approach to creating the XY-model, presented here.
However, in this regime, the evolution becomes more complex and deviates from expectations, despite the low occupancy of phonons (Fig.\ref{fig:Floquet}(d)).

For larger systems, numerical simulations are intractable.
Experimentally finding an appropriate $N_f$ is also challenging, as this would require thermometry of all the motional modes at multiple times in the evolution.
Further, a good $N_f$ with low Trotter error might require reducing the overall magnitude of the $J$-couplings pushing the dynamical timescale beyond the coherence time.
A good $N_f$ value simply may not even exist due to the number of motional modes participating in mediating the spin-spin couplings.
Additionally, creating arbitrary interaction graphs \cite{Korenblit_2012} may not be possible because optimal $N_f$ for different detunings (SDFs) will be different. 
The two SDF-approach of creating the XY interaction, presented here, provides a viable way to scale up quantum simulations.

In summary, we have demonstrated tunable long-range XY-type couplings ($J_{ij}^x$\XX + $J_{ij}^y$\YY) by the parallel application of two spin-dependent forces on the same motional modes.
Our approach allows for analog quantum simulations of the XY and anisotropic XY models, as the effective Hamiltonian (Eq. (\ref{eq_main_result})) is valid in continuous time.
This opens possibilities to study ground state order of frustrated XY-type models, in large 2D arrays of ions \cite{Britton_2012_Engineered}, and in principle on programmable lattice geometries \cite{Korenblit_2012_Quantum, Teoh_2020_Machine}, to investigate exotic quantum phases, such as spin liquids \cite{Varney_2011_Kaleidoscope}.
Further, evolution under the XY Hamiltonian can be interspersed with single spin quantum gates in analog-digital hybrid quantum simulations \cite{Rajabi_2019_Dynamical} to investigate dynamical phase transitions, Hamiltonian quenches, and quantum transport.
Our demonstration of parallel SDFs on the same motional modes can further be extended to simulate XYZ-type Hamiltonians by adding a $\sigma_z-$SDF (readily implemented using the light-shift gate schemes \cite{Baldwin_2021_High-fidelity} or \cite{bazawan_2023_synthesizing}). 

\begin{acknowledgments}
We thank Manoj Joshi for insightful discussions about floquet engineering techniques and practical issues while applying them to ions. We thank Jingwen Zhu for helping us on the experimental setup. We acknowledge financial support from the Natural Sciences and Engineering Research Council of Canada Discovery (RGPIN-2018-05250) program, Ontario Early Researcher Award, Canada First Research Excellence Fund (CFREF), New Frontiers in Research Fund (NFRF), University of Waterloo, and Innovation, Science and Economic Development Canada (ISED).
\end{acknowledgments}
\bibliographystyle{apsrev.bst}
\bibliography{apssamp}

\appendix
\onecolumngrid
\section{Distinguishing between XX and XY Hamiltonian (2 ions)}
If we start with the Hamiltonian $H = J^x_{12} \sigma_x^1 \sigma_x^2 + J^y_{12} \sigma_y^1 \sigma_y^2$ and the evolution operator U such that $U(\tau) = \mathrm{exp}(-iH\tau)$.
It is easy to show that:
\begin{itemize}
    \item U \downdown = $ \cos{\left(\tau \left(J^x_{12} - J^y_{12}\right) \right)}$ \downdown  - $  i \sin{\left(\tau \left(J^x_{12} - J^y_{12}\right) \right)}$ \upup
    \item U \upup = $ \cos{\left(\tau \left(J^x_{12} - J^y_{12}\right) \right)}$ \upup   - $ i \sin{\left(\tau \left(J^x_{12} - J^y_{12}\right) \right)}$ \downdown
    \item U \downup = $ \cos{\left(\tau \left(J^x_{12} + J^y_{12}\right) \right)}$ \downup  
     - $ i \sin{\left(\tau \left(J^x_{12} + J^y_{12}\right) \right)}$ \updown
     \item U \updown = $ \cos{\left(\tau \left(J^x_{12} + J^y_{12}\right) \right)}$ \updown  
     - $ i \sin{\left(\tau \left(J^x_{12} + J^y_{12}\right) \right)}$ \downup
\end{itemize}
For the case of the XY Hamiltonian, we have that $J^x_{12} = J^y_{12}$.
When initialized in \downdown, we do not expect to see oscillations between ( \downdown, \upup ).
But when initialized in \downup, we expect to see oscillations between ( \downup , \updown ) at a frequency of ($J^x_{12} + J^y_{12}$).

\section{Detailed derivation of constraint in Eq. (\ref{eq_main_result_constraint}) }
After RWA to Eq. (\ref{eq_two_SDF}) and with $\delta_{xm} = \omega_m - \mu_1$ and $\delta_{ym} = \omega_m - \mu_2$ we have, 
\begin{equation}
\begin{split}
    &H_{x} = \sum_{i,m}  \eta_{im}\Omega_i^x  ( a_m e^{-i (\delta_{xm} t  + \psi_x)} + h.c.) \sigma_x^i,  \\
    &H_{y} =  \sum_{i,m}  \eta_{im}\Omega_i^y  ( a_m e^{-i (\delta_{ym} t + \psi_y)} + h.c.) \sigma_y^i
\end{split}
\end{equation}
The first two terms in the exponent of (\ref{eq_2SDF_U}) come from calculations already described in \cite{Kim_2009_Entanglement}. 
The last two terms come from the cross commutators,
\begin{equation*}
\begin{split}
[H_x(t_1),H_y(t_2)] + [H_y(t_1),H_x(t_2)] &= [\sum_{i,m}  \eta_{im}\Omega_i^x  ( a_m e^{-i (\delta_{xm} t_1  + \psi_x)} + h.c.) \sigma_x^i, 
    \sum_{j,n}  \eta_{jn}\Omega_j^y  ( a_n e^{-i (\delta_{yn} t_2 + \psi_y)} + h.c.) \sigma_y^i ] \\ 
    & + [\sum_{i,m}  \eta_{im}\Omega_i^y  ( a_m e^{-i (\delta_{ym} t_1  + \psi_y)} + h.c.) \sigma_y^i, 
    \sum_{j,n}  \eta_{jn}\Omega_j^x  ( a_n e^{-i (\delta_{xn} t_2 + \psi_x)} + h.c.) \sigma_x^i ]
\end{split}
\end{equation*}
\begin{equation}
\begin{split}
    &=  \sum_{i,m,n} \eta_{im} \eta_{in} \Omega_i^x \Omega_i^y ( a_m e^{-i (\delta_{xm} t_1  + \psi_x)} + h.c.)
    ( a_n e^{-i (\delta_{yn} t_2  + \psi_y)} + h.c.) [\sigma_x^i,\sigma_y^i] \\
    &+  \sum_{i,m,n} \eta_{im} \eta_{in} \Omega_i^y \Omega_i^x ( a_m e^{-i (\delta_{ym} t_1  + \psi_y)} + h.c.)
    ( a_n e^{-i (\delta_{xn} t_2  + \psi_x)} + h.c.) [\sigma_y^i,\sigma_x^i]   \\
    &+ \sum_{i,j,m}  \eta_{im} \eta_{jm} \Omega_i^x \Omega_j^y  [( a_m e^{-i (\delta_{xm} t_1  + \psi_x)} + h.c.), ( a_m e^{-i (\delta_{ym} t_2  + \psi_y)} + h.c.)] \sigma_x^i \sigma_y^j \\
    &+ \sum_{i,j,m} \eta_{im} \eta_{jm} \Omega_i^y \Omega_j^x  [( a_m e^{-i (\delta_{ym} t_1  + \psi_y)} + h.c.), ( a_m e^{-i (\delta_{xm} t_2  + \psi_x)} + h.c.)]  \sigma_y^i \sigma_x^j \\
    &= 2 i \sum_{i,m,n} \eta_{im} \eta_{in} \Omega_i^x \Omega_i^y ( a_m e^{-i (\delta_{xm} t_1  + \psi_x)} + h.c.) 
    ( a_n e^{-i (\delta_{yn} t_2  + \psi_y)} + h.c.)  \sigma_z^i \\
    &- 2 i \sum_{i,m,n} \eta_{im} \eta_{in} \Omega_i^x \Omega_i^y ( a_m e^{-i (\delta_{ym} t_1  + \psi_y)} + h.c.) 
    ( a_n e^{-i (\delta_{xn} t_2  + \psi_x)} + h.c.)  \sigma_z^i \\
    &+ 2i \sum_{i,j,m} \eta_{im} \eta_{jm} \Omega_i^x \Omega_j^y \sigma_x^i \sigma_y^j  \big(\sin( \delta_{ym} t_2 - \delta_{xm} t_1 + \psi_y - \psi_x) - \sin( \delta_{ym} t_1 - \delta_{xm} t_2 - \psi_y + \psi_x) \big) 
\end{split}
\end{equation}
Using the above expressions ( and performing the integrals in Eq. (\ref{eq_2SDF_U}) we find that For $\mu_1 \neq \mu_2$,  there is no secular term in $\Lambda_{ij}(\tau)$ and $\hat{\zeta}_i(\tau)$, however, the oscillatory term in $\Lambda_{ij}(\tau)$ could get unbounded due to the presence of ($\delta_{xm} - \delta_{ym}$) in the denominator. 
 Let us define $\Tilde{\Lambda}_{ij}$ to be the coefficient of the largest oscillatory term in $\Lambda_{ij}(\tau)$.
 It can then be shown that
\begin{equation}
    |\Tilde{\Lambda}_{ij}| \leq \frac{4 \times \max( |J^x_{ij}| , |J^y_{ij}| )}{ |\mu_1 - \mu_2| }.
\end{equation}
When Eq. (\ref{eq_main_result_constraint}) is satisfied,  $\Tilde{\Lambda}_{ij}$ remains small and the contribution of the cross terms in Eq. (\ref{eq_2SDF_U}) becomes negligible and the XY-type couplings remain in the effective Hamiltonian.

\section{Implications of Eq. (\ref{eq_main_result_constraint}) }
In the text we mention that the Eq. (\ref{eq_main_result_constraint}) in text is weaker constraint than the the one imposed by applying the SDFs in the weak regime to keep the phonon excitation low.
For 2 ions and a SDF at $\mu$, we have $J_{12} \approx \frac{(\eta \Omega)^2}{\mu - \omega_m}$ and the probability of phonon excitation, $P_{ph} \approx \frac{(\eta \Omega)^2}{(\mu - \omega_m)^2}$.
To keep $P_{ph}$ low we apply the drives in the slow regime i.e $\eta \Omega \ll |\mu - \omega_m|$.
In general, for a chain of ions we require that 
\begin{equation}
    |\mu - \omega_m| \gg \eta_{im}\Omega_i
\end{equation}
For state of the art quantum simulators, typical values of radial trap frequencies for ($\omega_m$ here) are about 5 MHz, and while operating in the slow regime, typical values of Jmax are close to 100Hz even for a single SDF in the weak regime.
The spacing between the radial modes is on the order of 10s of kHz and hence there is``plenty of room to park" $\mu_1$ and $\mu_2$ close to each other while not getting too close to other modes.
In Fig (3) \ref{fig:Jij} we conservatively chose $|\mu_1 - \mu_2|$  = 3 kHz however $|\mu_1 - \mu_2| = 10 \times \max(J^{x(y})$ is enough.
Hence Eq. \ref{eq_main_result_constraint} does not limit the scale of the couplings any more than the limit imposed by being in the weak SDF limit.

\section{Floquet Analysis}
\begin{figure}[h]
\includegraphics[width=\linewidth]{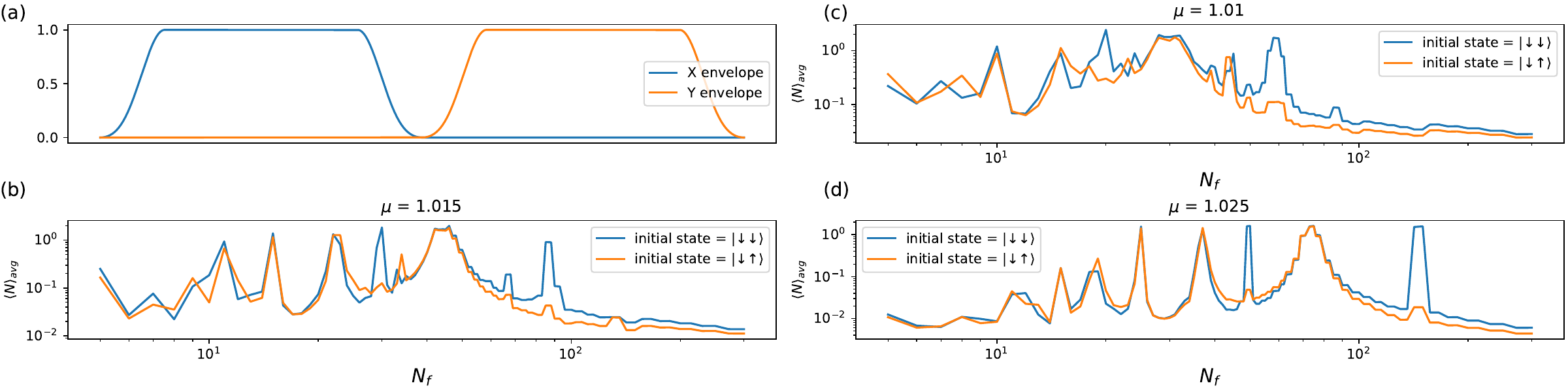}
\caption{Motional Excitation from periodic drives}
\label{fig:appendixFigFloquet}
\end{figure}
\textbf{(a)} For the simulation results shown Figure \ref{fig:Floquet}, we use a pulse sequence that effectively switches the spin phase the SDF at the floquet frequency.
We used a single motional mode to mediate the spin-spin interactions with $\omega_m = 2\pi \times 1$, and $\eta \Omega = 2\pi \times 0.003$  (see Eq. \ref{eq_MS_interaction_Hamiltonian}) for both XX and YY pulses.
The envelopes describe how the evolution of the Rabi frequency in each pulse.
The rising and falling edges are pulse shaped using a Blackman envelope function with the rise and fall times set to 20\% of the pulse time respectively.
\textbf{(b), (c) and (d)} In the manuscript we showed time averaged phonon occupation (and time evolution at specific values of $N_f$) for $\mu = \pi \times 1.02$
Here we show the average phonon occupation for other values of $\mu$ and find that the at the same Rabi frequencies and evolution time.

\section{Using motional modes in orthogonal directions to mediate interactions}
For mediating parallel XX and YY interactions on an ion chain, one possible alternative could also be to selectively address motional modes using frequency of the SDF.
Optical setup: Raman beams have overlap in  both X and Y modes.
Selective addressing of the modes using only frequency control is challenging because of the following reasons:
\begin{itemize}
    \item Maximum separation that can be achieved between the X and Y radial trap motional modes is limited  
    \item Adding ions in the chain makes the motional modes spread apart in frequency  
    \item Reducing the axial trap frequency to reduce phonon mode bandwidth leads to other challenges  
\end{itemize}

In the manuscript we parked our detunings very close to one motional mode to make sure that the dominant contribution to the both the XX and the YY couplings come from the same motional mode. This was done to confirm our theoretical prediction that indeed the same motional mode can mediate multiple interactions.

For power law type decay in couplings like the ones shown in Fig 3 of the manuscript, the detunings can go as far as 1 MHz for the power law exponent $\alpha \approx 2$.
If we restrict up to $\alpha \approx 1$, the detuning still has to be allowed to go to 100kHz.
Near detuned couplings, therefore have a very limited utility for useful quantum simulations.

Moreover, with more ions in the trap the additional motional modes along each trap axes spread apart in frequency.
For example, in our trap (see figure below),  the $\approx 200$ kHz spacing of the motional modes is not enough to separate them enough to satisfy the requirement of selective addressing even for just 10 ions.
Reducing the axial frequency to reduce the frequency bandwidth of modes in each direction has the added cost of making the ion chain spatial extent too large, and reduction in the overall scale of the J couplings because the COM and TILT modes in each direction get closer and lead to cancellations in the coupling strengths.
It might be tempting to think that the issue could be mitigated by simply making the two trap frequencies further separated.
However, skewing the trap can only be done to an extent since this brings the trap operation closer to the boundary of the stable trapping region.
\begin{figure}
    \centering
    \includegraphics[width=\linewidth]{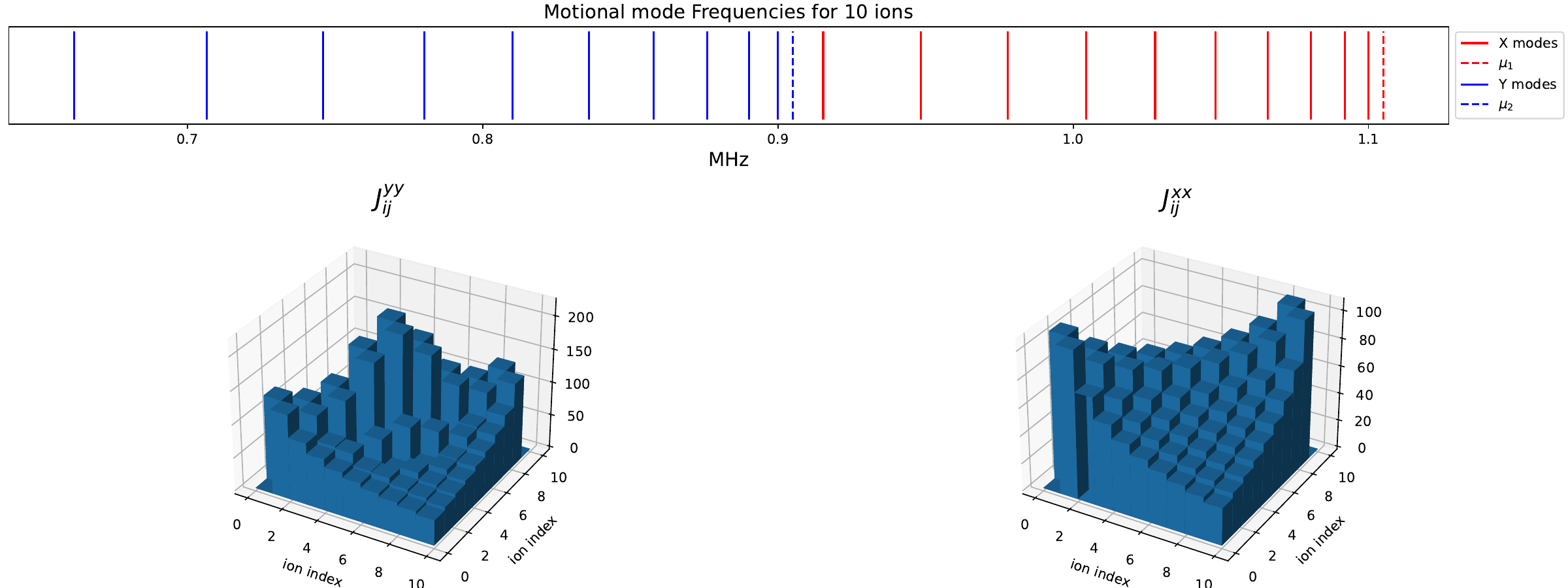}
    \caption{XY interactions mediated by separate motional modes}
Trap frequencies: $\omega_x = 2 \pi \times 1.1$, $\omega_y = 2 \pi \times 0.9$ MHz and $\omega_z = 2 \pi \times 132.7$ kHz.
The $\omega_z$ was chosen to the maximum value that allows the X and Y modes to be well separated.
This is because lower $\omega_z$ values lead to very long ion chains, very close proximity of the COM and TILT modes in both directions and large zero point spread of the ion positional wavefunction.
These effects make it difficult to address the ions optically and also lead to overall lower J couplings when $\omega_z$ is low.
\end{figure}

\end{document}